\documentclass[12pt]{article}

\usepackage[utf8]{inputenc}
\usepackage{geometry}
\usepackage{graphicx}
\usepackage{wrapfig}
\usepackage{cantarell}
\usepackage{tensor}
\usepackage{amsmath}
\usepackage{amssymb}
\usepackage{mathtools}
\usepackage{mathrsfs}
\usepackage{bbm}
\usepackage{slashed}
\usepackage[toc,page]{appendix}
\usepackage[backend=bibtex, style=phys]{biblatex}
\usepackage[english]{babel}
\usepackage[font=scriptsize, labelfont=bf]{caption}
\usepackage{subcaption}
\usepackage{csquotes}
\usepackage{float}
\usepackage[linktocpage]{hyperref}
\usepackage{comment}
\usepackage{authblk}
    
\graphicspath{{./figures}{./figures/eps}{./figures/png}{./figures/pdf}}

\DeclareGraphicsExtensions{.eps,.png}

\title{Positive Energy Warp Drive from Hidden Geometric Structures}

\author[1]{Shaun D. B. Fell\thanks{shfell@student.ethz.ch}}
\author[1]{Lavinia Heisenberg\thanks{laviniah@phys.ethz.ch}}
\affil[1]{Institute for Theoretical Physics, ETH Zürich, Wolfgang-Pauli-Strasse 27, 8093, Zürich, Switzerland}

\begin{document}
\renewcommand{\abstractname}{\vspace{-\baselineskip}}
 \begin{titlepage}

  \maketitle
  
    \begin{abstract}
    Warp drives in Einstein's general theory of relativity provide a unique mechanism for manned interstellar travel. It is well-known that the classical superluminal soliton spacetimes require negative energy densities, likely sourced by quantum processes of the uncertainty principle. It has even been claimed by few that negative energy densities are a requirement of superluminal motion. However, recent studies suggest this may not be the case. A general decomposition of the defining variables and the corresponding decomposition of the Eulerian energy are studied. A geometrical interpretation of the Eulerian energy is found, shedding new light on superluminal solitons generated by realistic energy distributions. With this new interpretation, it becomes a relatively simple matter to generate solitonic configurations, within a certain subclass, that respect the positive energy constraint. 
Using this newfound interpretation, a superluminal solitonic spacetime is 
presented that possesses positive semi-definite energy. A modest numerical analysis is carried out on a set of example configurations, finding total energy requirements four orders of magnitude smaller than the solar mass. Extraordinarily, the example configurations are generated by purely positive energy densities, a tremendous improvement on the classical configurations. The geometrical interpretation of the Eulerian energy thus opens new doors to generating realistic warp fields for laboratory study and potential future manned interstellar travel.
    \end{abstract}

\end{titlepage}

\newpage

\section{Introduction}

\paragraph{}
    The prospect of travelling to other far away astronomical systems has 
been a tantalizing thought for many millenia, however its bleak reality has only been realized in the past couple centuries. The vastness of space 
is the greatest barrier to interstellar travel. The closest planetary system to Earth is Alpha Centauri at a distance of 4.3 light-years \cite{wilkinson_new_2012}. Using the fastest man-made object, the Parker Space Probe, it would take approximately 65 centuries to reach Alpha Centauri \cite{apl_mission_nodate}. This is many times longer than the average human life span. Fortunately, there may exist other methods to manned intersystem travel using more fundamental theoretical processes. 
    
\paragraph{}
    It is well-known that an empty, flat, and non-dynamical spacetime does not permit massive particles from exceeding the speed of light. However, there have been cases in more complex systems of particles exceeding the local speed of light, such as highly energetic massive radiation in a medium giving rise to Cherenkov radiation \cite{Cherenkov_1934}. In some instances, even spacetime itself can travel faster than the vacuum speed of light. This is the case during the inflationary or dark energy phases of the universe. Such distortions of spacetime manifest as objects travelling faster than light relative to distant observers. This effect can be observed today by measuring the recession speed of distant galaxies and noting they are receeding faster than light away from the earth \cite{freedman_carnegie_chicago_2019, davis_expanding_2004}. 

\paragraph{}
    The dynamical nature of spacetime giving rise to superluminal relative speeds may be able to be exploited for interstellar travel. The field equations of general relativity (GR) permit inertial observers to be transported at superluminal rates relative to distance observers. The first solution possessing such features was the Alcubierre spacetime \cite{alcubierre_warp_1994}. This solution describes a region of spacetime in which an observer inside a ``spacetime bubble'' could be transported superluminally relative to observers outside the region. This is due to the contraction of spacetime in front of the central observer and a corresponding expansion behind. In theory, this solution permits interstellar travel of a payload on arbitrarily short timescales. 
    
\paragraph{}
    Unfortunately, this solution possesses numerous pathologies that would prevent its construction using realistic matter fields. Invoking quantum inequality restrictions enforced on the stress-energy tensor, as derived by \cite{ford_averaged_1995}, one finds this solution requires an enormous amount of negative energy, several orders 
of magnitude more than all the energy content in the observable universe \cite{pfenning_unphysical_1997}. 
Even if one appealed to quantum processes to generate the negative energies, the shear quantity requirement yields this configuration highly unphysical. To generate the energy densities required by the Alcubierre configuration using the Casimir force, the length scales would have to be millions of times smaller than the diameter of a proton. Additionally, the spacetime hosts horizons that would render the soliton uncontrollable \cite{hiscock_quantum_1997}. Other pathologies include subluminal weak energy violations and Planck-scale dimensions \cite{lobo_fundamental_2004, pfenning_unphysical_1997}.
    
\paragraph{}
    A number of other configurations have been constructed, each with varying goals and properties. The Natario spacetime is another example of a superluminal spacetime \cite{natario_warp_2002}. This solution, by construction, has trivial expansion. The advantage of this configuration is that it preserves volumes, as opposed to the Alcubierre spacetime which dramatically alters them. However, this configuration still comes with many of the same problems that the Alcubierre spacetime hosts. The Broeck configuration aims to reduce the massive amount of negative energy by altering 
the geometry of the warping region \cite{broeck_warp_1999}. They reduce the surface area of the "bubble" while maintaining the internal volume. This drastically reduces the 
total energy requirements from universal-scale to solar mass-scale. Nonetheless, negative energies are still required in order to generate all previously mentioned spacetimes. In recent years, theoretical research has hinted at possible superluminal configurations generated by purely positive energy densities \cite{lentz_breaking_2020}. This is accomplished by analyzing constraint equations enforced by GR and finding relationships that permit superluminal motion generated by positive energies. Other studies have claimed to generate superluminal motion from positive energies by first choosing a particular form of the stress-energy tensor and solving the Einstein equations in the Alcubierre spacetime \cite{2021GReGr..53...23S,2021EPJC...81..133S,2020EPJC...80..786S}. These studies appear to be in contrast to the claimed proof that superluminal motion requires negative energy densities \cite{olum_superluminal_1998}.
    
\paragraph{}
    Modified theories of gravity provide alternate descriptions of the gravitational force that could yield more physically realistic superluminal 
solitons \cite{Heisenberg:2018vsk}. One relevant example is the Debenedictis et al. configuration developed in the Einstein-Cartan theory \cite{debenedictis_energy_2018, cartan_comptes_1922, cartan_manifolds_1986}. In this theory of gravity, the torsion tensor no longer vanishes, resulting in new couplings between the spacetime geometry and matter fields (a similar setup can be constructed in terms of non-metricity \cite{BeltranJimenez:2017tkd, trinitygravity, BeltranJimenez:2018vdo}). An interesting feature of the Einstein-Cartan theory is that deviations from GR only manifest in the presence of matter. In this theory, the spin of the matter fields appears in the field equations. Debenedictis et al. exploited this coupling to generate the Alcubierre configuration using positive energy densities only. This, however, solves only one pathology of the Alcubierre configuration. It 
doesn't solve the horizon problems, causality problems, etc. Nonetheless, 
it illustrates the possible advantages of utilizing alternate theories of 
gravity to construct more physically realistic superluminal solitonic spacetimes. 
    
\paragraph{}
    With the presence of negative energy densities being one of the main pitfalls of the classical solitonic configurations, the aim of this work is to analyze the energy equation at the classical level in the context of superluminal spacetimes. The main question of this work is could a compact superluminal soliton described by the classical equations of general relativity be generated only using positive energy densities? Moreover, the geometric nature of the Eulerian energy within a certain subclass of spacetime configurations is investigated, providing insight into how and why the classical configurations necessitate negative energies. A few spacetimes hosting superluminal motion generated by positive energies will also be presented, along with some of the subtleties of the Eulerian energy.
    
\paragraph{}
    This paper is then divided into three main sections. Section 2 will introduce the relevant theoretical tools and their relevance to the superluminal solitonic spacetimes. Section 3 will present the results of this study and its significance. Finally, section 4 will conclude with a summary and discussion on potential future work.

\section{Theory}

\subsection{Background}
\paragraph{}
    Henceforth, the conventions of \cite{gourgoulhon_31_2012} will be adopted, namely Latin indices are assumed to run from 1 to 3, Greek indices run from 0 to 3, the metric signature is assumed to be $(-,+,+,+)$, and their sign conventions are adopted. The starting point is the decomposition of the Einstein equations in the (3+1)-decomposition. Spacetime is assumed to be globally hyperbolic. This implies the topology is $\Omega \times \mathbb{R}$, where $\Omega$ is a Cauchy surface, since the flow of timelike vector fields intersecting the Cauchy surface provides a homeomorphism between the spacetime $\mathscr{M}$ and $\Omega \times \mathbb{R}$. This allows the spacetime to be foliated by a family of spacelike hypersurfaces, whose extrinsic geometry is described by a normal vector. Since the normal vector is a timelike vector, 
it can be thought of as the 4-velocity of an observer. This observer is called an \emph{Eulerian observer}.
    
\paragraph{}
    The internal geometry of the hypersurfaces is described by the 3-metric $\gamma_{ij}$, which is induced by the pull-back between the spacetime 
$\mathscr{M}$ and hypersurface $\Omega$. This metric has an associated Levi-Civita connection and, hence, a corresponding Riemann curvature tensor. This curvature tensor describes the intrinsic curvature of the hypersurface. Moreover, the evolution of the 3-metric along the normal direction gives the extrinsic curvature of the hypersurface as an embedded manifold,
    \begin{equation} \label{def:extrinsiccurvature}
        K_{i j} = - \frac{1}{2} \mathcal{L}_n \gamma_{ij}.
    \end{equation}
    Formally, the extrinsic curvature is defined as the inner product between a tangent vector to the hypersurface and the shape operator. This definition is equivalent to Eq. \eqref{def:extrinsiccurvature}. 
    
\paragraph{}
    The Gauss-Codazzi relations are important relations between the Einstein tensor and the intrinsic and extrinsic geometry of the hypersurfaces. They represent 
various projections of the curvature tensor of the full spacetime onto the hypersurface and its normal. Their contracted forms can be compactly written using the Einstein tensor as
    \begin{align} \label{eq:gausscodazzi}
        G_{\mu \nu}n^{\mu}n^{\nu} &= \frac{1}{2} \left( \tensor[^\Omega]{\mathscr{R}}{} + \mathscr{K}^2 - K_{ij}K^{ij} \right)\\
        \nonumber
        G_{\mu \nu}n^{\mu}\tensor{\gamma}{^{\nu}_{\alpha}} &= D_{\alpha}\mathscr{K} - D_{\mu}\tensor{K}{^{\mu}_{\alpha}},
    \end{align}
    where $G_{\mu \nu} \coloneqq \tensor[^4]{R}{_{\mu \nu}} - \frac{1}{2} 
\tensor[^4]{\mathscr{R}}{} g_{\mu \nu}$ is the four-dimensional Einstein tensor, $n^{\mu}$ 
is the hypersurface normal vector, $\tensor[^\Omega]{\mathscr{R}}{}$ is the Ricci scalar calculated from the 3-metric, $\mathscr{K}$ is the trace of the extrinsic curvature, and $D_{\alpha}$ is the Levi-Civita connection associated to the 3-metric. Importantly, these relations represent the bilateral projection, resp. mixed projection, of the Einstein tensor onto 
the hypersurface normal, resp. hypersurface normal and the hypersurface itself. Adopting the definitions of \cite{wald_general_1984}, the extrinsic curvature trace represents the expansion of a congruence of geodesics passing through the hypersurface,
    \[
        \theta \coloneqq K^{ij}\gamma_{ij} = \mathscr{K}.
    \]

     Using co-ordinates adapted to the foliation, the normal vector can be decomposed into a scalar and vector component as $n^{\mu} = (\frac{1}{N}, - \frac{\vec{N}}{N})$. The scalar component $N(t,x,y,z)$, called the 
\emph{lapse function}, describes in a sense the spacing of the leaves of the foliation, while the vector component $\vec{N}(t,x,y,z)$, called the \emph{shift vector}, describes the tangential shift of the co-ordinate system between neighboring leaves. The full spacetime metric can then be decomposed in terms of the lapse function, shift vector, and 3-metric as
    \[
        g_{\mu \nu} = 
        \begin{bmatrix}
            -N^2 + \vec{N} \cdot \vec{N} & N_j \\
            N_i & \gamma_{ij}.
        \end{bmatrix}
    \]
    Importantly, a shift vector field satisfying $\vec{N} \cdot \vec{N} > 
N^2$ is called \emph{superluminal} due to the fact $g(\partial_t, \partial_t) >0$.
\paragraph{}
    In the presence of matter fields, the stress-energy tensor also admits a decomposition in terms of the foliation variables. The complete bilateral projection of the stress-energy tensor onto the hypersurface normal represents the energy density of the matter field, while the complete bilateral projection onto the hypersurface is the mechanical stress. The mixed projection, once onto the hypersurface and once onto the normal, represents the momentum density (or energy flux) of the field.
    \begin{align} \label{def:stressenergyprojection}
        \text{ Energy Density: }& \rho = \mathscr{T}_{\mu \nu}n^{\mu}n^{\nu}\\\nonumber
        \text{ Momentum Density: }& p_{\sigma} = -\mathscr{T}_{\mu \nu}n^{\mu}\tensor{\gamma}{^{\nu}_{\sigma}}\\\nonumber
        \text{ Stress Tensor: }& S_{\alpha \beta} = \mathscr{T}_{\mu \nu}\tensor{\gamma}{^{\mu}_{\alpha}}\tensor{\gamma}{^{\nu}_{\beta}} 
    \end{align}
    Using the foliation-adapted co-ordinates and the contracted Gauss-Codazzi relations, the Einstein equations, $G_{\mu \nu} = \frac{8 \pi G_N}{c^4} \mathscr{T}_{\mu \nu}$, decompose under the foliation as
    \begin{align} \label{eq:decomposedeinsteinequation}
        &\left(\partial_t - \mathcal{L}_{\vec{N}}\right) \gamma_{ij} = -2 N K_{ij}\\\nonumber
        &\left(\partial_t - \mathcal{L}_{\vec{N}} \right)K_{ij} = -D_i D_j N + N \Big( \tensor[^{\Omega}]{R}{_{ij}} + \mathscr{K} K_{ij} - 2 K_{ik} \tensor{K}{^k_j} + 4 \pi \big( (\mathcal{S} - \rho) \gamma_{ij} - 2 S_{ij}\big) \Big)\\\nonumber
        &\tensor[^{\Omega}]{\mathscr{R}}{} + \mathscr{K}^2 - K_{ij}K^{ij} 
= 16 \pi \rho\\\nonumber
        &D_j \tensor{K}{^{j}_{i}} - D_i \mathscr{K} = 8 \pi p_i,
    \end{align}
    where $S_{ij}$ is the mechanical stress tensor and $\mathcal{S}$ its trace. The third and fourth equations are called the \emph{Hamiltonian constraint} and \emph{momentum constraint}, respectively.
    
\subsection{Weak Energy Condition}
\paragraph{}
    Investigating the physicality of superluminal spacetime solitons begins with studying the energy conditions enforced on the spacetime. One of the weakest energy conditions one can enforce, aptly named the weak energy condition (WEC), states that
    \[
       \text{WEC:  }  \;\; \mathscr{T}_{\mu \nu} t^{\mu} t^{\nu} \geq 0, \;\;\; \; \;\forall \text{  timelike  } t^{\mu}.
    \]
    Intuitively, this is the statement that the local energy density measured by any observer must be non-negative.
    This implies the Eulerian energy density, calculated from the Hamiltonian constraint, must be positive semi-definite. 

\paragraph{}
    The WEC is violated in the Alcubierre configuration. The chosen ansatz is $N=1$, $\gamma_{ij} = \delta_{ij}$, and $\vec{N} = (-v_s f(r_s),0,0)$ with $r_s^2 = x^2 + y^2 + z^2$. The Eulerian energy density is then easily calculated, using Eq. \eqref{def:extrinsiccurvature} and the Hamiltonian constraint, to be
    \[
        \rho = - \frac{1}{8 \pi} \frac{v_s^2 \rho^2}{4 r_s^2} \left(\frac{df}{dr_s} \right)^2,
    \]
    where $v_s$ is the velocity of the warp bubble and $\rho^2 = y^2 + z^2$ the radial cylindrical co-ordinate.
    Indeed, the fact that the weakest energy condition is violated by such a simple example of a superluminal soliton does not bode well for superluminal travel. Note that even at subluminal speeds, when $0<v_s<1$, the energy density is still negative. Similar results hold for the other classical superluminal spacetimes, and indeed it has been claimed that a couple of the classical configurations are equivalent, related only by a co-ordinate transformation \cite{bobrick_introducing_2021}. Moreover, it is no surprise the expansion-less Natario configuration violates the WEC. From the 
Hamiltonian constraint, if the expansion is zero, the energy density reads $\rho = - K_{ij}K^{ij}$, where intrinsic flatness was assumed. It follows that any configuration attempting to satisfy the WEC must contain a sufficiently large magnitude of expansion, at every spacetime point where 
the extrinsic curvature is non-vanishing, so that the Eulerian energy density is non-negative.

\section{Results}
\paragraph{}
Presented in this section is a class of solitonic spacetimes generated by 
positive energy densities, alongside a new geometric interpretation of the Eulerian energy distribution. The considered class of spacetimes permit 
superluminal transportation of an inertial observer. Several comparisons of the presented configurations can be drawn to previous constructions in 
the literature. The most widely used ansatz will also be employed here, namely the shift vector field will be used to fully describe the spacetime. Moreover, potential fields will be introduced to further analyze the constraining equations, similar to the analysis in \cite{lentz_breaking_2020}, albeit without enforcing extra constraints on the metric variables. The presented work appears to be in good agreement with the study performed in \cite{bobrick_introducing_2021}, namely the non-truncation of 
the gravitational field leads to positive semi-definite energies, axisymmetric configurations can lead to interesting solitonic configurations, and modifying the geometry of the soliton can lead to a reduction of the energy requirements. 

\paragraph{}
There are several advantages to the presented configurations compared to those previously presented in the literature. The foremost advantage is the absence of negative energies, in contrast to the Alcubierre, Natario, and Van Den Broeck configurations among others. The presented solitonic configurations are also tuneable, which is a first of its kind. This means the central observer can have a variable speed, relative to distant observers, by simply altering the geometry of the Eulerian energy distribution while maintaining the total energy. The speed is based upon the geometry of the energy distribution and 
not the total energy content, making it highly desirable for interstellar 
transportation. The configurations are also highly modifiable due to the geometric interpretation offered by the Hamiltonian constraint. This will 
greatly aid any work seeking to minimize the energy requirements.

\subsection{Generalities}
\paragraph{}
    In order to explore the feasibility of constructing a positive-energy 
spacetime soliton, the Eulerian energy is investigated in a suitably general ansatz. The shift vector field is taken to be time-independent with no other constraints placed on it. Moreover, the lapse function is taken to be unity, and the induced metric is the identity. The Hamiltonian constraint then reduces to
    \begin{align} \label{eq:WECinansatz}
        16 \pi \rho =& 2 \bigg( \partial_x N_x \partial_y N_y + \partial_x N_x \partial_z N_z + \partial_y N_y \partial_z N_z \bigg)\\
        \nonumber
        &- \frac{1}{2} \bigg( ( \partial_z N_y + \partial_y N_z)^2 + ( \partial_y N_x + \partial_x N_y)^2 + ( \partial_z N_x + \partial_x N_z)^2 \bigg)
    \end{align}
    Some observations are already in order. The first term in the above equation has an indeterminate sign and is quadratic in derivatives of different components. The second term is negative semi-definite and involves the square of linear combinations of derivatives. This explains why some of the previously proposed solitonic configurations exhibit negative energies. Their configurations depend only on a single component of the shift vector. For example, the Alcubierre ansatz reduces the above equation to
    \[
        16 \pi \rho = - \frac{1}{2} \bigg( (\partial_y N_x)^2 + (\partial_z N_x)^2 \bigg).
    \]
    In order to satisfy the WEC, any ansatz must contain at least two non-zero components of the shift vector field.
    
\paragraph{}
    Eq. \eqref{eq:WECinansatz} does not provide any further information regarding its positivity. One must dive deeper into the internal structure 
of the equation to investigate its properties. This can be accomplished by considering a general decomposition of the shift vector field. There are a number of decompositions of vector fields, but the most commonly employed, and the one utilized here, is the Helmholtz decomposition
    \begin{equation}\label{def:Helmholtz}
        \vec{N}(x,y,z) = \vec{\partial}\phi(x,y,z) + \vec{\omega}(x,y,z),
    \end{equation}
    where $\phi$ is a scalar field and $\vec{\omega}$ is a solenoidal field satisfying $\vec{\nabla} \cdot \vec{\omega}=0$. The metric tensor, in line element form, then reads
    \begin{align} \label{eq:METRICLINEELEMENT}
        ds^2 = -dt^2 + &\left(dx + \left(\partial_x \phi(x,y,z) + \omega_x(x,y,z)\right)dt\right)^2\\\nonumber
         +&\left(dy + \left(\partial_y \phi(x,y,z) + \omega_y(x,y,z)\right)dt\right)^2\\\nonumber
         +&\left(dz + \left(\partial_z \phi(x,y,z) + \omega_z(x,y,z)\right)dt\right)^2
    \end{align}
    
\paragraph{}
    After tedious algebra, the Hamiltonian constraint can be shown to reduce to
    \begin{equation}\label{eq:WECinhelmholtz}
        16 \pi \rho = 2*\bigg( h_1 + h_2 + h_3 \bigg) - 2 \bigg \langle 
\mathcal{J}, \mathcal{H} \bigg \rangle_F - \bigg \langle \mathcal{J}, \mathcal{J} \bigg \rangle_F + \frac{1}{2} | \vec{\nabla} \times \vec{\omega}|^2,
    \end{equation}
    where 
    \[
    \mathcal{H}(x,y,z) = 
        \begin{bmatrix}
            \partial_x^2 \phi(x,y,z) & \partial_x \partial_y \phi(x,y,z) & \partial_x \partial_z \phi(x,y,z)\\
            \partial_y \partial_x \phi(x,y,z) & \partial_y^2 \phi(x,y,z) & \partial_y \partial_z \phi(x,y,z)\\
            \partial_z \partial_x \phi(x,y,z) & \partial_z \partial_y \phi(x,y,z) & \partial_z^2 \phi(x,y,z)
            \end{bmatrix}
    \]
    is the Hessian matrix of $\phi(x,y,z)$,
    \[
    \mathcal{J}(x,y,z) = 
        \begin{bmatrix}
            \partial_x \omega_x(x,y,z) & \partial_y \omega_x(x,y,z) & \partial_z \omega_x(x,y,z)\\
            \partial_x \omega_y(x,y,z) & \partial_y \omega_y(x,y,z) & \partial_z \omega_y(x,y,z)\\
            \partial_x \omega_z (x,y,z)& \partial_y \omega_z(x,y,z) & \partial_z \omega_z(x,y,z)
        \end{bmatrix}
    \]
    is the Jacobian matrix of $\omega(x,y,z)$,
    \begin{align*}
        h_1(x,y,z) &= \left(\partial_y^2 \phi(x,y,z)\right) \left(\partial_z^2 \phi(x,y,z)\right) - \left( \partial_y \partial_z \phi(x,y,z) \right)^2\\
        h_2(x,y,z) &= \left(\partial_x^2 \phi(x,y,z)\right) \left(\partial_z^2 \phi(x,y,z)\right) - \left( \partial_x \partial_z \phi(x,y,z) \right)^2\\
        h_3(x,y,z) &= \left(\partial_x^2 \phi(x,y,z)\right) \left(\partial_y^2 \phi(x,y,z)\right) - \left( \partial_x \partial_y \phi(x,y,z) \right)^2\\
    \end{align*}
    are the three second-order principal minors of $\mathcal{H}$, $\vec{\nabla}\times \vec{\omega}$ is the curl of the solenoidal field, and the sesquilinear form $\langle \cdot, \cdot \rangle_F$ is the Frobenius product defined as
    \[
     \left \langle A, B \right \rangle_F = \sum_{i,j} A_{ij}B_{ij}.
    \]
    Importantly, it is a positive-definite form, implying the third term in 
Eq. \eqref{eq:WECinhelmholtz} is negative definite. The momentum constraint admits a simple decomposition,
    \[
        16 \pi \vec{p} = \nabla ^2 \vec{\omega} = - \vec{\nabla} \times \vec{\nabla} \times \vec{\omega}.
    \]
    Notably, this implies irrotational fields will always enforce a vanishing momentum density. Henceforth, explicit co-ordinate dependence will be dropped unless required for clarity, e.g. $h_1(x,y,z) \rightarrow h_1$, etc.
    
\paragraph{}
    The key result of this decomposition is the geometrical interpretation of the energy density in the purely irrotational sector. For shift vector fields described by a single scalar field, $\vec{N} = \vec{\partial} 
\phi$, the constraint equations in the (3+1)-decomposition of the Einstein equations reduce to
    \begin{align} \label{eq:WECinirrotationalhelmholtz}
        8 \pi \rho &= h_1 + h_2 + h_3\\\nonumber
        \vec{p} &= 0.
    \end{align}
    The WEC then enforces the constraint
    \[
        h_1 + h_2 + h_3 \geq 0.
    \]
    The scalar component of the shift vector field contributes a geometric term to the Eulerian energy density. The second-order principal minors describe the curvature of the scalar field in the two-dimensional subspaces. Although it is not directly related to the convexity of the entire function, convexity of the two-dimensional subspaces of the scalar field $\phi$ determines the positivity of the Eulerian energy density. Convexity (concavity) of the entire function is indeed a sufficient condition, but it is not necessary. As long as one of the principal minors is sufficiently 
large, the other two may be negative while still satisfying the positive energy constraint.
    
\paragraph{}
    Although not evident from Eq. \eqref{eq:WECinhelmholtz}, its a simple 
task to show that purely solenoidal shift vector fields have negative semi-definite energy. This follows directly from the Hamiltonian constraint in Eq. \eqref{eq:decomposedeinsteinequation}, noting that $\mathscr{K} = 
\vec{\nabla} \cdot \vec{N}$. If the shift vector field were purely solenoidal, then $\vec{\nabla} \cdot \vec{N} = 0$, implying $\rho = - \frac{1}{16 \pi} K_{ij}K^{ij} \leq 0$, as is the case for the Natario spacetime.

\subsection{Positive Energy Spacetime Soliton} \label{sec:positivenergywarpdrive}
\begin{figure}
        \begin{subfigure}{0.5\textwidth}
            \centering
            \includegraphics[width=0.75\linewidth]{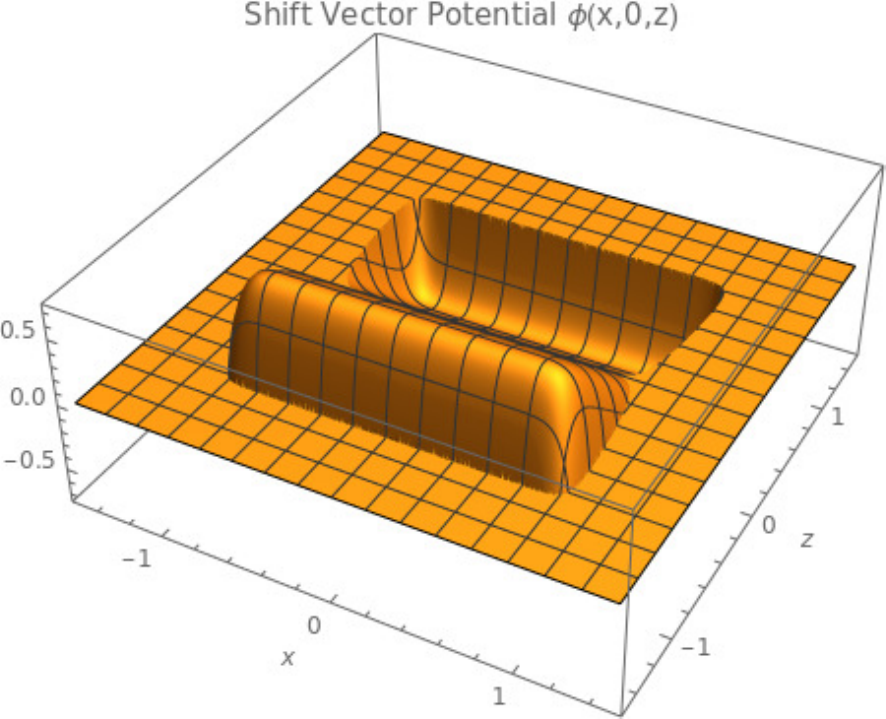}
            \caption{}
            \label{fig:PWslicedfunc}
        \end{subfigure}
        \begin{subfigure}{0.5\textwidth}
            \centering
            \includegraphics[width=0.8\linewidth]{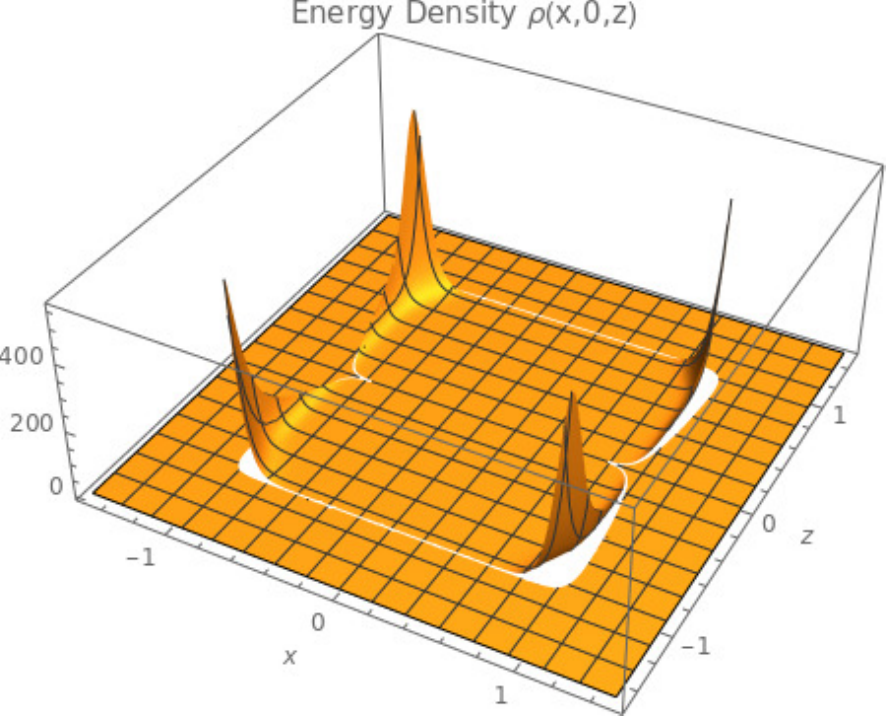}
            \caption{}
            \label{fig:PWslicedenergy}
        \end{subfigure}
        \begin{subfigure}{0.5\textwidth}
            \centering
            \includegraphics[width=0.85\linewidth]{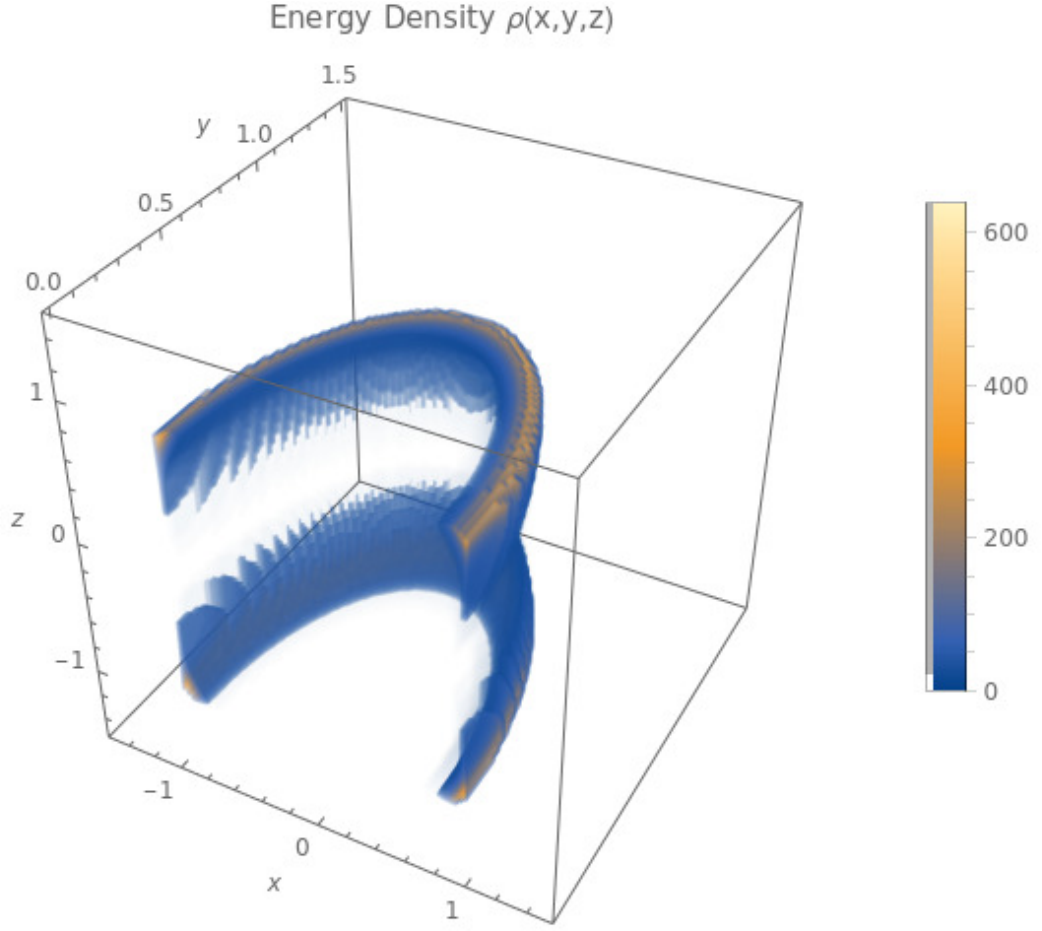}
            \caption{}
            \label{fig:PWfullenergy}
        \end{subfigure}
        \begin{subfigure}{0.5\textwidth}
            \centering
            \includegraphics[width=0.75\linewidth]{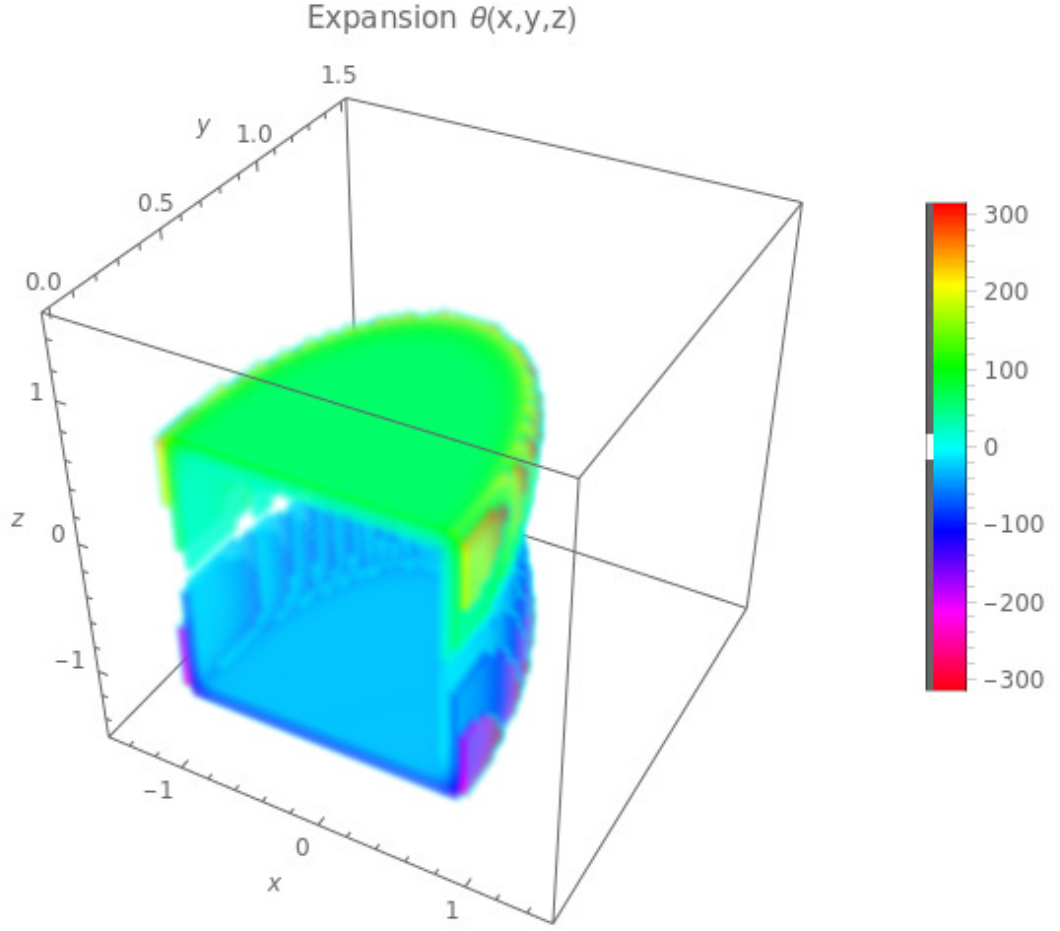}
            \caption{}
            \label{fig:PWfullexpan}
        \end{subfigure}
        \begin{subfigure}{0.5\textwidth}
            \centering
            \includegraphics[width=0.7\linewidth]{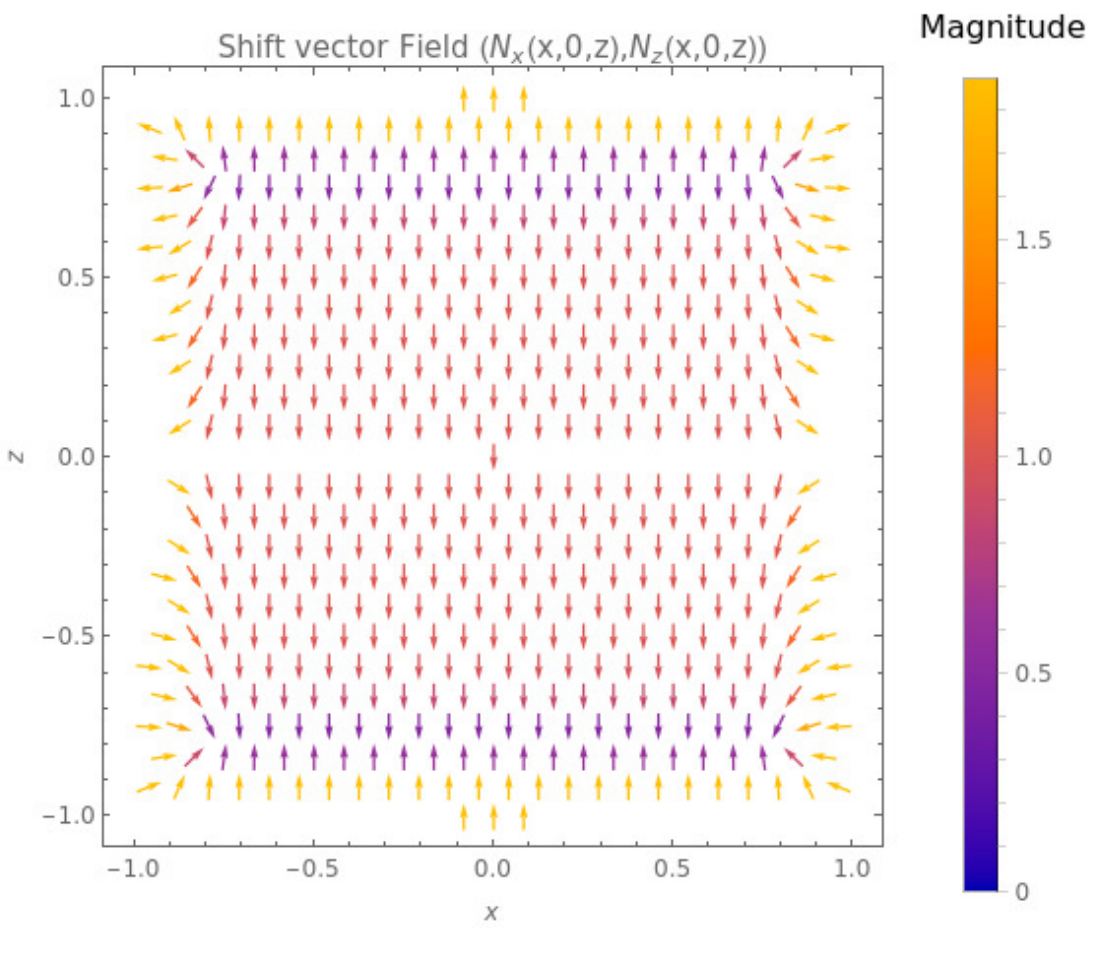}
            \caption{}
            \label{fig:PWslicedshift}
        \end{subfigure}
        \begin{subfigure}{0.5\textwidth}
            \centering
            \includegraphics[width=0.83\linewidth]{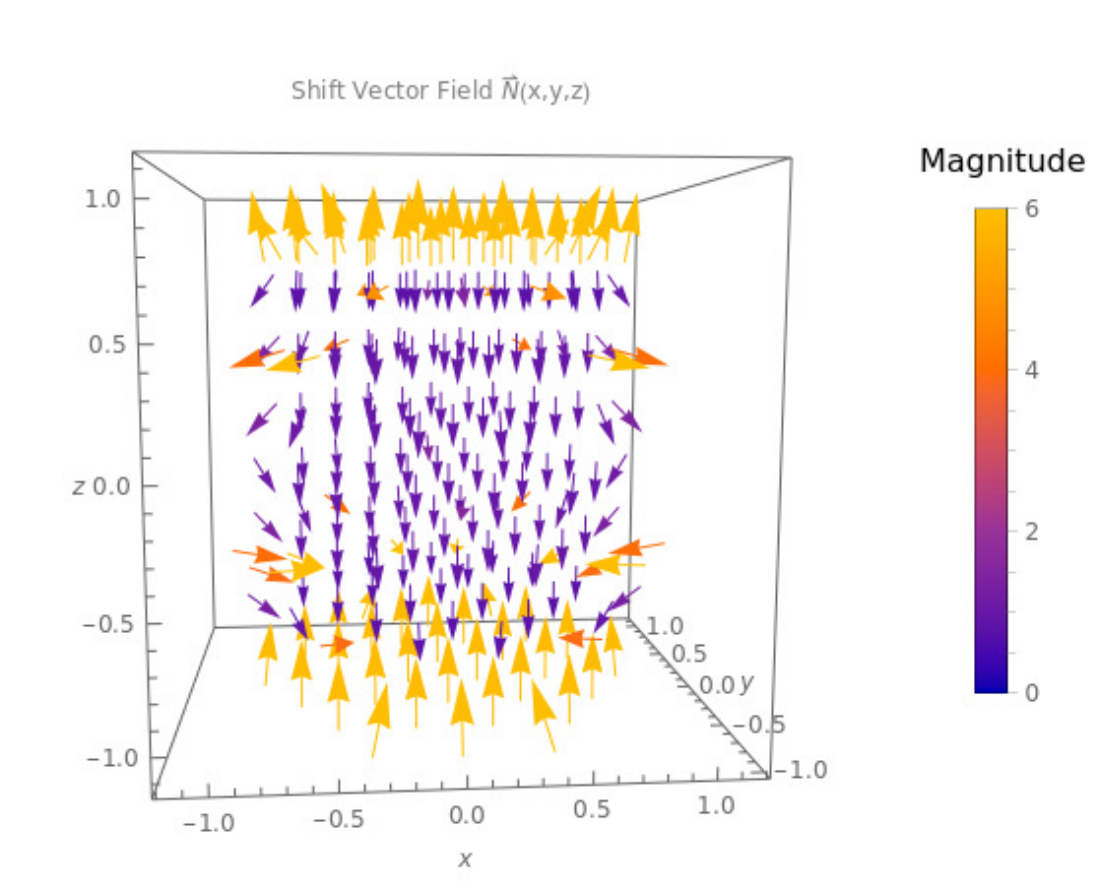}
            \caption{}
            \label{fig:PWShift}
        \end{subfigure}
        \caption{Configuration described by a piecewise-defined shift vector potential. Figure \ref{fig:PWslicedenergy} shows the energy density of the configuration in the $y=0$ plane. Note the large discontinuities. 
Figure \ref{fig:PWfullenergy} shows a 3D density plot of the energy density. Opacity and color are a proxy for magnitude. The magnitude is maximal 
on the outer edge of the toroids and minimal inside. Figure \ref{fig:PWfullexpan} shows a density plot of the volume expansion. The magnitude of the expansion is maximal on the outer face of the toroid and minimal on the inside. Positive expansion is occurring in the positive z region, while 
contraction is occurring in the negative z region. Figure \ref{fig:PWslicedshift} is the $y=0$ slice of the two shift vector components $(N_x, N_z)$. In a neighborhood of the origin, the shift vector field is almost purely in the negative z-direction (towards the negative expansion). }
    \label{fig:PWSoliton}
    \end{figure}

\paragraph{}
    Using the geometrical interpretation offered by Eq. \eqref{eq:WECinirrotationalhelmholtz}, it is a simple matter to construct, using geometric intuition alone, a positive energy solitonic spacetime. Assume the shift vector field is purely irrotational and let
    \begin{equation}
        \phi(x,y,z) = -1.3z + 1.3
        \begin{cases}
        - \big(\sqrt{x^2 + y^2}\big)^{10} - z^4, & -\big(\sqrt{x^2 + y^2}\big)^{10} - z^4>z\\
        \big(\sqrt{x^2 + y^2}\big)^{10} + z^4, &  \big(\sqrt{x^2 + y^2}\big)^{10} + z^4<z\\
        z, & \text{else}
        \end{cases}.
    \end{equation}
    The shift vector field is then calculated from $\vec{N} = \vec{\partial} \phi$. The scalar field $\phi$ fully describes the spacetime, including the energy density and expansion. See Fig. (\ref{fig:PWSoliton}). The direction of travel for this configuration is in the negative z-direction. The total integrated expansion is zero and the magnitude of the shift vector in 
the central region around $(x,y,z)=(0,0,0)$ is 1.3, which is superluminal. Moreover, the 
energy density approaches zero in the central region, a perfect location for a vehicle. Additionally, the shift vector field portrays the same directionality as the Alcubierre configuration, namely the shift vector field points away from expanding regions and towards contracting regions. It is important to note that the scalar field is not $C^1$-differentiable.
\paragraph{}
    One could construct several variations of the above configuration. The key trait is that its energy is purely positive. One could seek to minimize the energy density while maintaining superluminality of the central region by altering the geometry of the scalar field. However, the discontinuities present in the configuration could pose a large problem when attempting to construct such a configuration.

\paragraph{}
    One must be careful when dealing with piecewise functions in spacetime. Cusps, corners, and other discontinuities aren't physical. As suggested by the geometrical nature of Eq. \eqref{eq:WECinirrotationalhelmholtz}, 
negative energies will manifest if the scalar field $\phi$ has strongly-curved saddle points in any of the three two-dimensional subspaces. One could ``hide'' the saddle points in the discontinuities of the function, such as a corner. Any physically realistic field should be sufficiently smooth, since any physical construction will inevitably possess a certain fall-off in the otherwise geometrical corner. Additionally, the length scale of such a discontinuity is infinitesimally small, way beyond the regime 
of classical gravity. One would need a description of the small-scale nature of spacetime to even begin to describe configurations containing scales of this order. 
    
\begin{figure}
    \begin{subfigure}{0.5\textwidth}
        \centering
        \includegraphics[width=0.8\linewidth]{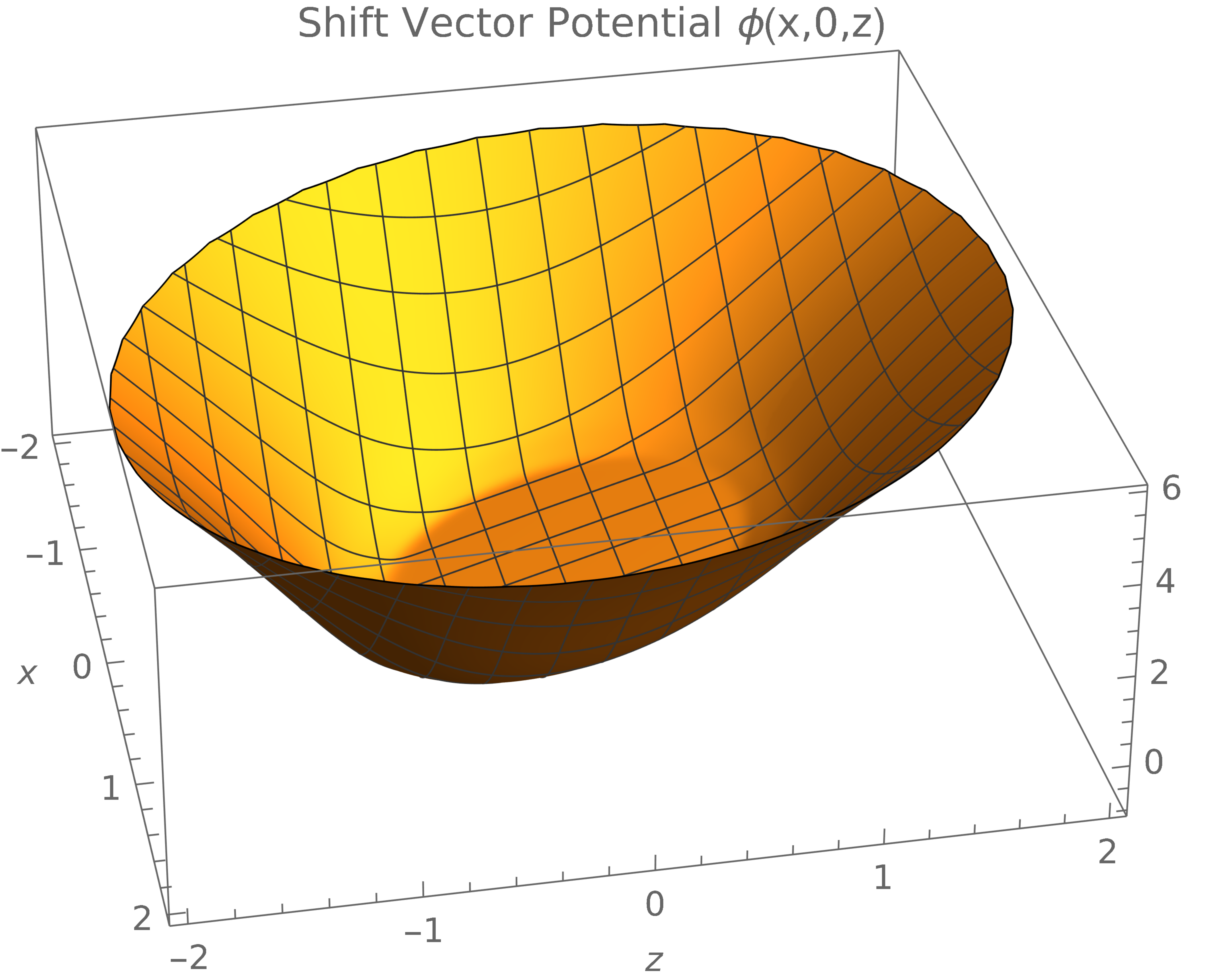}
        \caption{}
        \label{fig:IrrotThreeslicedfunc}
    \end{subfigure}
    \begin{subfigure}{0.5\textwidth}
        \centering
        \includegraphics[width=0.8\linewidth]{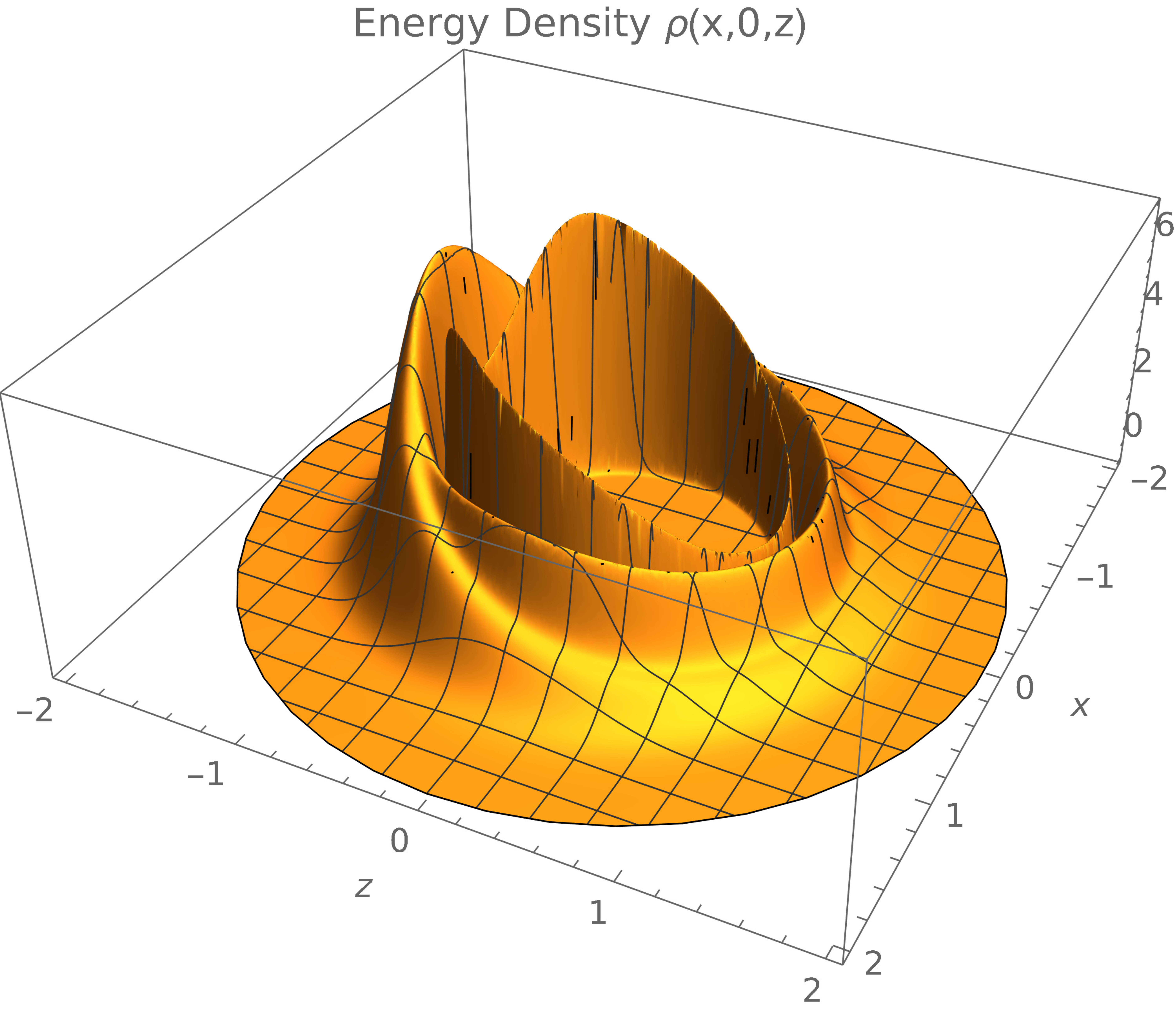}
        \caption{}
        \label{fig:IrrotThreeSlicedEnergy}
    \end{subfigure}
    \begin{subfigure}{0.5\textwidth}
        \centering
        \includegraphics[width=0.8\linewidth]{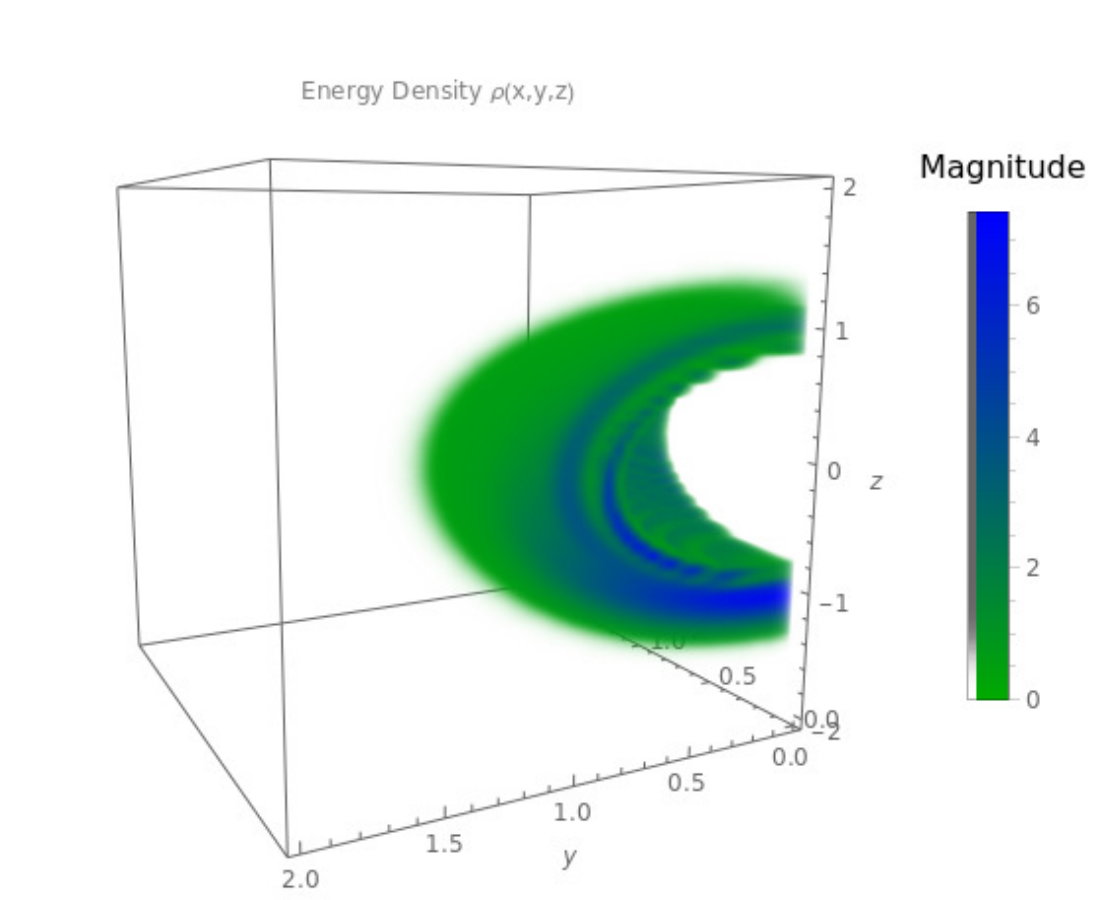}
        \caption{}
        \label{fig:IrrotThreeFullEnergy}
    \end{subfigure}
    \begin{subfigure}{0.5\textwidth}
        \centering
        \includegraphics[width=0.8\linewidth]{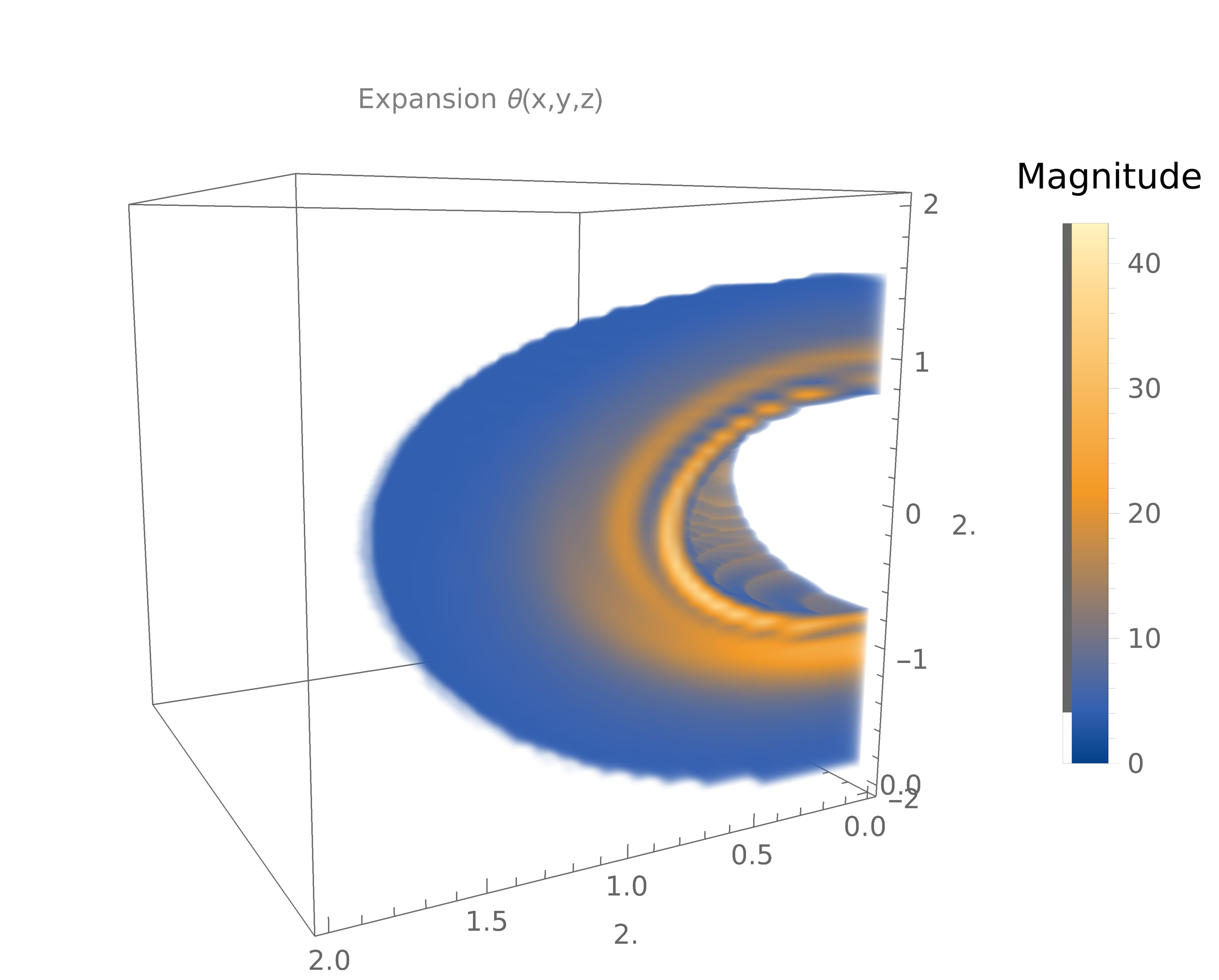}
        \caption{}
        \label{fig:IrrotThreeFullExpan}
    \end{subfigure}
    \begin{subfigure}{\textwidth}
        \centering
        \includegraphics[width=0.4\linewidth]{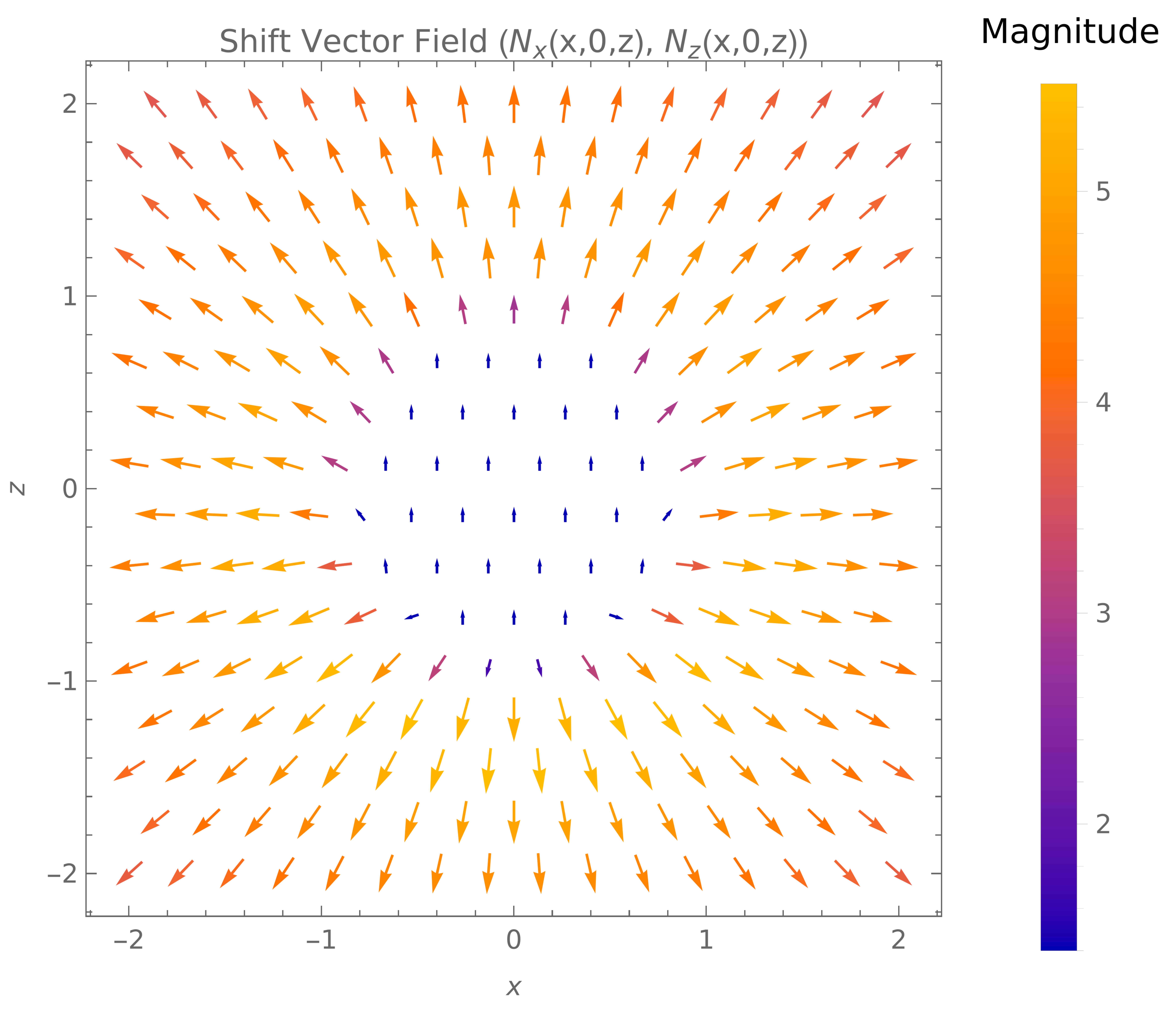}
        \caption{}
        \label{fig:IrrotThreeFullShift}
    \end{subfigure}
    \caption{Configuration defined from an at least $C^2$-differentiable shift vector potential, shown in Figure \ref{fig:IrrotThreeslicedfunc}. The energy density, Figure \ref{fig:IrrotThreeSlicedEnergy} ($y=0$ slice) and Figure \ref{fig:IrrotThreeFullEnergy} (3D density plot), is highly localized. It has a finite integral over the three-dimensional leaves of the foliation. The expansion, Figure \ref{fig:IrrotThreeFullExpan} (3D density plot), is everywhere non-negative, with greater expansion in the negative z region. Opacity and color in the density plots serve as a proxy for magnitude. The shift vector field, Figure \ref{fig:IrrotThreeFullShift}, is highly radial and asymptotically Minkowskian in the exterior region and highly linear in the interior with a central magnitude of $\sim 1.26$, i.e. superluminal.}
    \label{fig:irrotsuperluminal}
\end{figure}

\paragraph{}
    In the interest of constructing a physically interesting superluminal 
configuration, an additional smoothness condition is enforced on the spacetime. The shift vector potential field $\phi$ is required to be at least 
$C^2$-differentiable. This forces the resulting energy density distribution to be continuous, largely avoiding issues associated to many piecewise 
configurations. One such example of a scalar field satisfying this condition is

\begin{align} \label{def:irrotational1potential}
     \phi(x,y,z) =& \frac{V}{m(x,y,z)+n(x,y,z)} \left( \sigma \left( e^{-\frac{\left(r - \frac{(x^2+y^2+z^2)^\Pi}{m(x,y,z)} \right)^2}{\sigma}}m(x,y,z)n(x,y,z) \right. \right. \\\nonumber
     & \left. \left.+ e^{-\frac{\left(r + \frac{(x^2+y^2+z^2)^\Pi}{n(x,y,z)} \right)^2}{\sigma}}m(x,y,z)n(x,y,z) - e^{-\frac{r^2}{\sigma}}\left(m(x,y,z) + n(x,y,z)\right) \right) \right.\\\nonumber
     & \left.+ \sqrt{\sigma \pi} \left(-\left(\left(m(x,y,z) + n(x,y,z)\right)r*\text{Erf}\left[\frac{r}{\sqrt{\sigma}}\right] \right) \right. \right.\\\nonumber
     & \left. \left. + n(x,y,z)*\left(m(x,y,z)*r-(x^2+y^2+z^2)^\Pi\right)*\text{Erf}\left[\frac{r-\frac{(x^2+y^2+z^2)^\Pi}{m(x,y,z)}}{\sqrt{\sigma}}\right] \right.\right. \\\nonumber
     & \left.\left. + m(x,y,z)*\left(n(x,y,z)*r-(x^2+y^2+z^2)^\Pi\right)*\text{Erf}\left[\frac{r+\frac{(x^2+y^2+z^2)^\Pi}{n(x,y,z)}}{\sqrt{\sigma}}\right] \right) \right),
\end{align}

    where $\Pi$, $r$, and $V$ are parameters to be specified, $\sigma$ is a Gaussian weight parameter, $m(x,y,z)$ and $n(x,y,z)$ are spacial functions used to set the central gradient, and $\text{Erf}[]$ is the error function. For certain choices of $m(x,y,z)$, $n(x,y,z)$, and $\Pi$, this is $C^2$ on the region $\mathbb{R}^3 - 0$ and approaches zero as $\vec{x} \rightarrow \vec{0}$. For example, the choice
    \begin{align*}
        \left(\Pi,r,V, \sigma\right) &=\left(\frac{1}{4}, 6, 10, 1\right)
    \end{align*}
    and suitable spherically symmetric choices for $m(x,y,z)$ and $n(x,y,z)$ results in a smooth function on $\mathbb{R}^3-0$ and approaches zero as $\vec{x} \rightarrow \vec{0}$. The energy density is everywhere positive and highly localized. Moreover, the total energy,
    \[
        E_{\text{total}} = \int dx dy dz \rho(x,y,z),
    \]
    is finite. The central region $|\vec{x}|<r$ has asymptotically vanishing shift vector. The only superluminal shift vector is in the exterior region $|\vec{x}|>r$, which approaches Minkowski space at asymptotic infinity. A simple modification to the spacial functions, $m(x,y,z)$ and $n(x,y,z)$, results in a configuration possessing superluminal shift in the interior region, Fig. (\ref{fig:irrotsuperluminal}). By letting $m(x,y,z)>n(x,y,z)$ in the $+z$ region and $m(x,y,z)<n(x,y,z)$ in the $-z$ region,  this results in the gradient of the $\phi$-field increasing in the region $|\vec{x}|<r$, generating a high and level shift. The direction of travel is now in the positive z-direction. Compare this to the spherically symmetric case for $m(x,y,z)$ and $n(x,y,z)$. The $\phi$-field becomes spherically symmetric, resulting in the energy density being uniformly distributed in a spherical shell around the central region $|\vec{x}|<r$. Modifying the central gradient causes the energy density distribution to be highly peaked in the $-z$ region. For the specific choice of parameters 
used in Fig. (\ref{fig:irrotsuperluminal}), the central shift magnitude is 1.26, i.e. superluminal. 
    
\paragraph{}
    This could serve as a method of generating acceleration without modifying the total energy on the hypersurface. One could start with a finite energy distribution highly concentrated and uniformly distributed in a spherical shell around a spacecraft. The energy density is then increased behind the spacecraft and decreased in front, without destroying or creating any additional energy. This increases the shift vector magnitude in the region where the spacecraft is located, transporting it to non-zero speeds relative to faraway observers. If the energy density is sufficiently concentrated behind the central observer, the relative speed becomes superluminal. This then operates as a tuneable solitonic configuration capable of superluminal speeds. Whether the central vehicle can actually manipulate the energy density in the luminal or superluminal regimes remains in question. The spacetime might still possess some of the other pathologies associated to the classical solitonic configurations, 
such as the formation of horizons. Although, there are two important deviations from the classical configurations, namely the absence of negative Eulerian energies and the exterior region no longer behaving like Minkowski space. 
The exterior region of the warp bubble is now Schwarzschild-like and satisfies the prerequisites of Birkoff's theorem only in the case when $m(x,y,z)$ and $n(x,y,z)$ are spherically symmetric, which the spacetime in Fig. (\ref{fig:irrotsuperluminal}) does not. This non-truncation of the gravitational field outside the warping region is likely the reason why negative energies are absent from this configuration, in agreement with \cite{bobrick_introducing_2021}. Interestingly, by tuning the Gaussian weight, the energy density could be reduced without modifying the central 
gradient of the $\phi$-field. This could be used to avoid the formation of a black hole due to high energy densities. This configuration is then highly modifiable, thanks to the geometric interpretation of the Eulerian energy. Compared to the 
configuration presented in \cite{lentz_breaking_2020}, the shift vector potential in Eq. \eqref{def:irrotational1potential} is more general as it does not assume any relations between the shift vector components. Moreover, as there are no relations between the derivatives of the shift vector 
potential, the scalar field can be easily adjusted in many ways in order to 
optimize the physical construction of the configuration.
    
\paragraph{}
    With the Eulerian energy distribution in hand, the full WEC can now easily be investigated to study the nature of the energy distribution in other frames of reference. By analyzing the Lorentz-invariant eigenvalues of the stress-energy tensor calculated using Eq. \eqref{eq:METRICLINEELEMENT}, it turns out the WEC is violated in compact regions within the energy distribution. It follows from the Lorentz-invariant eigenvalue equation that the Eulerian energy, Eq. \eqref{eq:WECinirrotationalhelmholtz}, is the first eigenvalue (whose eigenvector is the normal vector) and thus satisfies the first constraint enforced by the WEC for a type-I stress-energy tensor, which the configuration in Fig. (\ref{fig:irrotsuperluminal}) appears to be \cite{poisson_2004,10.1093/ptep/ptaa009}. However, the principle momenta constraints is where the violations occur, namely, $\rho + p_i^{p}<0$ for some $i \in \{1,2,3\}$ in compact regions in the distribution. No amount of modification to the configuration could get rid of these WEC-violating regions. However, the WEC is not violated everywhere and if the configuration in \cite{lentz_breaking_2020} indeed satisfies the WEC, as claimed, then it may still be possible to satisfy the WEC in the presented configurations too, given sufficient modifications.

\paragraph{}
    The ansatz Eq. \eqref{eq:WECinansatz} contains an interesting property related to an O(3) operation. The dynamical equations Eq. \eqref{eq:decomposedeinsteinequation} are not invariant under the total spacial inversion $\vec{N} \rightarrow -\vec{N}$. Additionally, the expansion is not invariant either, since under this operation, $\mathscr{K} \rightarrow -\mathscr{K}$. However, since $\gamma_{ij}=\delta_{ij}$, the inversion is equivalent, according to the dynamical equations, to the temporal inversion $t \rightarrow -t$. As expected, the shift vector inversion is equivalent to an inversion of the foliation label, which can be interpreted as ``time running backwards.'' This means that the same static energy distribution can produce two different configurations, one in which spacetime possesses positive expansion and one with negative expansion. It makes sense, then, to consider the strong energy condition (SEC) in this case. The (geometric) SEC is the constraint $\tensor[^4]{R}{_{\mu \nu}}t^{\mu} t^{\nu} \geq 0$ for all timelike $t^{\mu}$. In the case of a purely irrotational shift vector field, this condition in the Eulerian frame takes the simple form
    \[
        \tensor[^4]{R}{_{\mu \nu}}n^{\mu}n^{\nu} = - \langle \mathcal{H}, \mathcal{H} \rangle_F - \vec{\nabla} \phi \cdot \vec{\nabla}( \nabla^2 
\phi).
    \]
    It is interesting that this equation is invariant under the inversion $\phi \rightarrow -\phi$. This means that if the spacetime possessing positive expansion everywhere violates the SEC, then so too will the spacetime obtained by $\phi \rightarrow -\phi$, which possesses negative expansion (contraction) everywhere. In fact, it follows by simple calculations that if
    \[
        \partial_t (\vec{\nabla} \cdot \vec{N}(t,x,y,z))= \partial_t \theta(t,x,y,z) = 0
    \]
    then $\tensor[^4]{R}{_{\mu \nu}}n^{\mu}n^{\nu}$ is invariant under $\vec{N} \rightarrow - \vec{N}$. This means that if the expansion is independent of co-ordinate time, then the geometric side of the SEC constraint in the Eulerian frame is invariant under the total inversion $\vec{N} \rightarrow - \vec{N}$. Evidently, the configurations in Fig. (\ref{fig:PWSoliton}) and Fig. (\ref{fig:irrotsuperluminal}) violate the SEC. 

\paragraph{}
    Even though the Eulerian momenta vanish, and thus satisfies the constraint enforced by the dominant energy condition in the Eulerian frame, the full dominant energy condition is also violated due to the fact $|\rho|-|p_i^{p}|<0$ for some $i\in{1,2,3}$ in compact regions in the distribution, where $p_i^{p}$ are the principal momenta lorentz-invariant eigenvalues.
    
\paragraph{}
    One of the main issues with warp drive spacetimes is trying to find possible stress-energy sources that generate the proposed spacetimes. Warp drive research typically starts with a given geometry and attempts to find what stress-energy distribution sources the geometry via the Einstein equations, as the study presented here does. This doesn't guarantee the resulting stress-energy source to be physical. Starting with a stress-energy source in a given geometry and finding what conditions must be respected for the source to be physical is still on-going research. An explicit study of this sort is out of the scope of this paper; however, some points can still be mentioned. The configuration proposed in \cite{lentz_breaking_2020} is based on the same ansatz studied in this paper. In principle, a similar model could be applied to the configurations presented here. Moreover, the sources discussed in \cite{2021EPJC...81..133S, 2021GReGr..53...23S, 2020EPJC...80..786S} could be applied to the configurations here as well. More complete statements on possible stress-energy sources would require a deeper analysis, which is reserved for future studies.

\subsection{Chronological Protection} \label{sec:chronological}
When studying superluminal motion, one inevitably runs into the problem of generating closed timelike curves (CTCs) and the associated chronological protection conjecture presented in \cite{hawking_chronology_1992}. As discussed, the configurations presented in Sec. \ref{sec:positivenergywarpdrive} admit superluminal inertial observers. This naturally leads to the 
question of whether these configurations can be used to generate CTCs or not. It may be that these configurations can be used to violate causality 
in the superluminal regime, however there are two possible arguments for reconciliation with the chronological protection conjecture. The first is 
that one of the preconditions of the presented configurations is that the 
spacetime is globally hyperbolic. This means that the leaves of the foliation are Cauchy surfaces and hence CTCs cannot form. The second is that as the energy distribution is being manipulated from the spherically symmetric distribution into the axi-symmetric distribution (Fig. \ref{fig:irrotsuperluminal}), horizons form preventing manipulation of the energy distribution past the speed of light barrier. In other words, horizon formations prevent the configurations from transporting an inertial observer from the subluminal regime to the superluminal regime. This agrees with the discussion in \cite{bobrick_introducing_2021}. More work will need to be done to analyze the formation of horizons and the evolution from the subluminal to superluminal regimes in the presented configurations.

\subsection{Numerics} \label{sec:numerics}
\paragraph{}
To get a sense of the magnitude of the energy densities in the previous section, Newton's constant and the speed of light are reinserted into the Hamiltonian constraint. With this definition of the units, the energy density becomes
\[
\rho(x,y,z) =\frac{c^4}{16 \pi G} \left[ 2*\bigg( h_1 + h_2 + h_3 \bigg) - 2 \bigg \langle \mathcal{J}, \mathcal{H} \bigg \rangle_F - \bigg \langle \mathcal{J}, \mathcal{J} \bigg \rangle_F + \frac{1}{2} | \vec{\nabla} 
\times \vec{\omega}|^2 \right].
\]
 
\paragraph{}
For the case of an irrotational field, this simplifies, as before, to
\[
 \rho(x,y,z) = \frac{c^4}{8 \pi G} (h_1 + h_2 + h_3).
\]

\paragraph{}
The maximum value of the energy density for the configuration shown in Fig. \ref{fig:irrotsuperluminal}, as measured by the Eulerian observer, is found to be 
\[
\rho_{\text{max}} \approx 3.2*10^{26} \text{   $\frac{kg}{m^3}$}.
\]
This is an astronomical energy density concentrated in a very small region. More than likely, any attempt to construct this specific configuration 
will form a black hole. The total energy present on any initial hypersurface can be found to be approximately
\[
 E_{total} \approx 9.25*10^{43} \text{      $J$},
\]
which is still enormous, but quite small in astronomical terms. For comparison, the rest mass energy of the sun is approximately 
\[
 E_{\odot} \approx 1.78*10^{47} \text{        $J$}.
\]
Hence, the total energy required to form the configuration of Fig. (\ref{fig:irrotsuperluminal}) is four orders of magnitude smaller than the rest mass energy of the sun. This is a drastic improvement on the energy scales of the classical superluminal solitons. Moreover, thanks to the geometric interpretation of the Eulerian energy in the considered class of spacetimes, it would be a simple matter to drastically reduce the energy requirements, as discussed in Sec. \ref{sec:positivenergywarpdrive}, and thus avoid 
the formation of a black hole.

\section{Conclusion}    
\paragraph{}
    The results discussed in the previous sections provide great insight into the properties of the weak energy condition for superluminal solitonic spacetimes. The geometric interpretation of the Eulerian energy density opens new doors for constructing realistic superluminal spacetimes from 
feasible energy densities. Numerical techniques could be employed to investigate the nature of the irrotational sector of the Hamiltonian constraint, such as a Finite Element Method for solving the second order partial differential equation. The techniques available during this study could not solve the PDE, owing to its high non-linearity. More than likely, a new PDE solver would need to be constructed specifically for this differential equation. One could also perform other decompositions of the shift vector field to investigate other hidden properties. This may illuminate hidden interpretations that would allow easier construction of a shift vector field composed of irrotational and solenoidal components. 

\paragraph{}
    The energy densities discussed in Sec. \ref{sec:numerics} could easily be reduced to more manageable scales by minimizing the curvature of the shift vector potential. One could even utilize computational power to 
search for configurations that minimize the energy density while maximizing the central shift vector magnitude. 

\paragraph{}
    These results are extremely exciting. They shed new light on physically-realistic superluminal warp drives and brightens the future of manned interstellar travel.

\section*{Acknowledgements}
LH is supported by funding from the European Research Council (ERC) under the European Unions Horizon 2020 research and innovation programme grant agreement No 801781 and by the Swiss National Science Foundation grant 179740. 
 
\newpage
\phantomsection
\addcontentsline{toc}{section}{References}
\printbibliography
 
\end{document}